\documentclass[lettersize,journal]{IEEEtran}
\IEEEoverridecommandlockouts
\usepackage{amsmath,amsfonts}
\usepackage{booktabs}   
\usepackage{algorithmic}
\usepackage{algorithm}
\usepackage{array}
\usepackage[caption=false,font=normalsize,labelfont=sf,textfont=sf]{subfig}
\usepackage{textcomp}
\usepackage{xcolor}
\usepackage{stfloats}
\usepackage{url}
\usepackage{multirow}
\usepackage{verbatim}
\usepackage{graphicx}
\usepackage{cite}
\usepackage{hyperref}
\usepackage{fancyhdr}
\def\BibTeX{{\rm B\kern-.05em{\sc i\kern-.025em b}\kern-.08em
    T\kern-.1667em\lower.7ex\hbox{E}\kern-.125emX}}

\begin{document}

\title{Transient Triggering Grid-Forming Synchronization Control Under Voltage and Frequency Dips}

\author{Dewan Mahnaaz Mahmud,~\IEEEmembership{Student member,~IEEE,} {Vinu Thomas},~\IEEEmembership{Member,~IEEE,} {Bogdan Marinescu},~\IEEEmembership{Member,~IEEE}, {Micka\"{e}l Hilairet}

\thanks{Dewan Mahnaaz Mahmud, Vinu Thomas and Micka\"{e}l Hilairet are with Nantes Université, École Centrale Nantes, LS2N, UMR 6004, F-44000 Nantes, France  (email: dewan-mahnaaz.mahmud@ec-nantes.fr; vinu.thomas@ec-nantes.fr; mickael.hilairet@ec-nantes.fr).}

\thanks{Bogdan Marinescu is with École Centrale Nantes, 1 Rue de la Noë, 44000 Nantes, France 
(email: bogdan.marinescu@ec-nantes.fr).}


}



\maketitle
\thispagestyle{fancy}

\begin{abstract}
Grid-forming (GFM) inverters are gaining attention as a promising alternative for conventional synchronous generators in the modern power systems. Unlike conventional synchronous generators, GFM inverters have limited overcurrent capability that makes them vulnerable during large disturbances. During disturbances i.e., voltage and frequency dips, GFM inverters are pushed into current-limited operation to protect the semiconductor switches. This causes the internal angle of the GFM inverters to accelerate and lose synchronism with the rest of the grid. To address this limitation, this article proposes a transient triggering grid-forming (TTGFM) synchronization control to enhance the synchronization stability performance under voltage and frequency dips. This method uses two feedback signals; terminal voltage and the difference between unsaturated and saturated power to adjust the internal angle which is generated by power synchronization loop (PSL) of the GFM inverters. These two signals manipulates the internal reference angle generation that act as a virtual braking mechanism. This mechanism limits the angle acceleration during voltage and frequency dips without requiring an extra supervisory signal or parameters to tune. The proposed method is benchmarked against two state-of-the-art synchronization enhancement schemes and validated through high-fidelity electromagnetic transient (EMT) simulations with a grid dynamic equivalent (GDE) model in MATLAB/Simulink\textsuperscript{\textregistered}. An analytical framework is developed to derive the synchronization instability mechanism and the critical limits of the stability margins. Generalization of the GDE model further shows that the TTGFM control is not restricted to a single configuration but is applicable to any standard benchmark system.
\end{abstract}


\section{Introduction}
\IEEEPARstart{M}{odern} power systems are facing new control and stability challenges due to the increased penetration of power-electronic converters. To address these challenges, grid-forming (GFM) inverters have gained attention from both industry and academia \cite{ref1}. GFM inverters are identified as a potential replacement for conventional synchronous generators. That means, GFM inverters can be modeled as a controlled voltage source behind an internal impedance and can regulate voltage and frequency autonomously \cite{ref2}. 

Transient stability still remains a critical issue for stable operation of GFM inverters under large disturbances such as, e.g., voltage or frequency dips. Transient stability (also known as synchronization stability) is defined as the capability of the inverter to maintain synchronism with the grid after these large disturbances \cite{ref3}. Under voltage or frequency dips, the inverter might be forced to operate in current-limited mode to protect the semiconductor switches from over-current. Once the current limiter is activated, the phase angle of the inverter begins to accelerate. This acceleration causes windup of the internal angle and could lead to a loss of synchronism with the rest of the grid \cite{ref4}. 

Synchronization stability problem under voltage \cite{ref5}, \cite{ref6}, \cite{ref7} and frequency dips \cite{ref8}, \cite{ref9}, \cite{ref10} have each been studied extensively in the literature in independent manners. In \cite{ref11} authors proposed an adaptive droop control method to enhance the synchronization performance of GFM under voltage dips. This method is indeed effective as the droop gain is reduced once the current exceeds the threshold and by the difference between the measured current and that threshold. But this method might not be effective under frequency dips, as small droop gain could lead to more aggressive frequency responses. Another method \cite{ref12} is to freeze the  frequency and angle reference to the pre-fault condition. A similar limitation also hold for this method: while it is effective for voltage dips, it is likely to be ineffective for frequency dips because of the mismatch between the phase angles. Other methods \cite{ref13}, \cite{ref14}, \cite{ref15}, \cite{ref16} are also validated against voltage dips, and performance during frequency dips remains questionable. On the other hand, authors in \cite{ref17} proposed a power setpoint modification scheme based on a proportional gain that is based on the inverter current threshold. This method is effective for frequency dips but might not be effective for grid voltage dips. Similarly, in \cite{ref18} authors proposed an power matching-based GFM control to address synchronization stability during frequency excursions. The performance during voltage dips remains questionable here also. 

The capability of GFM inverters to remain synchronized with the grid during voltage and frequency dips is a requirement by the grid operators \cite{ref19}. Yet, state-of-the-art control solutions have made progress in enhancing performance on either voltage or frequency dip individually; few of them address both requirements at the same time. Authors in \cite{ref20} proposed a power-angle based control scheme to address both frequency and voltage dips. But this method requires a phase-locked loop (PLL) which can be vulnerable to delays and weak grid conditions. In \cite{ref21}, a solution based on virtual active power is proposed to increase transient stability margin. However, this method is limited to the implementation of virtual impedance as a voltage control loop. Authors in \cite{ref22} proposed the concept of fictitious power to increase the robustness of GFM against large disturbances. But this method requires tunable virtual impedance. While these methodologies can effectively enhance synchronization performance of GFM inverters under voltage and frequency dips, it is still possible to further maximize this performance. Also, these methodologies require parameter tuning, PLL, virtual impedance as voltage control loop, or external supervisory signals. 

To address these limitations, this article proposes an alternative method named as transient triggering grid-forming synchronization (TTGFM) control to enhance the synchronization stability of the GFM inverters against voltage and frequency dips. This proposed method uses two feedback signals: terminal voltage and the difference between saturated and unsaturated power. These feedback signals are the input of the power synchronization loop (PSL) of the GFM inverters to act as a virtual braking system to reduce the angle acceleration during voltage and frequency dips. This proposed method is triggered only during these transient events and effectively limits the angle acceleration to enhance synchronization stability performance. The proposed method also does not require an extra parameter to tune or external supervisory signals or PLL. Also, unlike previous validation methods that assume an infinite grid with constant voltage or frequency, TTGFM control is validated with a dynamic grid equivalent (GDE) model. Therefore, considering grid dynamics, the effectiveness of the TTGFM control is analytically analyzed along with the synchronization instability mechanism and critical limits. Furthermore, the applicability of TTGFM control to any standard benchmark system is presented through the generalization of the GDE model. Finally, the proposed method is validated against two state-of-the-art transient stability enhancement schemes through high-fidelity electromagnetic transient (EMT) simulations in MATLAB/Simulink\textsuperscript{\textregistered}. In summary, this article:
\begin{enumerate}
    \item[1)] Proposes a TTGFM control that enhances synchronization stability of GFM inverters under voltage and frequency dips without requiring an extra parameter to tune or external supervisory signals or
    PLL.
    \item[2)] Develops an analytical framework for synchronization stability analysis considering grid dynamics to understand the instability mechanism and critical stability limits.
    \item[3)] Demonstrates that the proposed TTGFM control and analytical framework is not restricted to a single grid configuration but applicable to any standard benchmark or real systems (e.g., IEEE 9, 14 or 39 bus) by generalization of GDE model.
    \item[4)] Benchmarks the proposed method against two state-of-the-art synchronization enhancement schemes through high-fidelity EMT simulations in MATLAB/Simulink\textsuperscript{\textregistered}.
\end{enumerate}

The rest of the article is organized as follows: In section \ref{S2}, the proposed method is introduced. Simulation-based characterization is presented in \ref{S3}. Benchmarking of the proposed method with the other state-of-the-art methods is presented in section \ref{S4}. In section \ref{S5}, analytical framework is presented that includes: synchronization instability mechanism, theoretical analysis of the proposed control, critical stability limits and effect of line impedance variation. Section \ref{S6} presents the applicability of the proposed control in any standard benchmark system. Finally, section \ref{S7} concludes the article.  

\section{Transient Triggering Grid-Forming Control}
\label{S2}

\begin{figure*}[t!]
\centering
\includegraphics[width=0.9\linewidth]{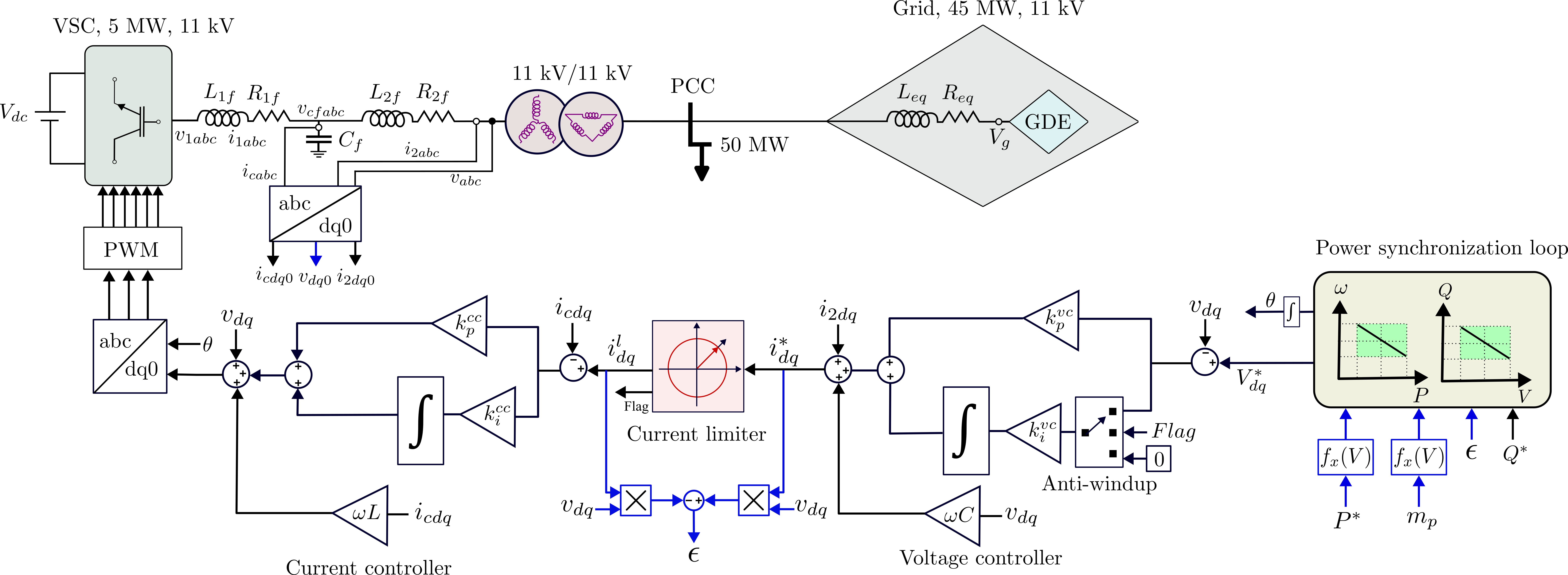}
\caption{System architecture with proposed TTGFM control.}
\label{fig:Sys_TTGFM}
\end{figure*}

Synchronization of GFM inverters with rest of the grid is achieved through PSL. Typically there are three types of PSL: droop control, virtual synchronous machine and virtual oscillator. Proposed TTGFM control is a modification of the conventional droop-based GFM control. Frequency regulation of TTGFM control is achieved by the expression given in \eqref{eq:TTGFM}. Here, $f(V)$ and $\epsilon(I)$ in \eqref{eq:PF} are the two piecewise functions. The function $f(V)$ scales the active power reference $P^{*}$ and droop gain $m_p$, while $\epsilon(I)$ is the difference between the unsaturated and saturated power that is fed back into the PSL.

\begin{equation}
\omega = \omega^{\ast} 
+ m_p f(V)
\Bigl(P^{\ast} f(V) - P - \epsilon(I)\Bigr)
\label{eq:TTGFM}
\end{equation}

\begin{subequations}
\label{eq:PF}
\begin{align}
f(V) &=
\begin{cases} 
1, & V > 0.9 \\
 V, & 0.5 < V \leq 0.9 \\
0, & V \leq 0.5
\end{cases}
\label{eq:AF} \\[8pt]
\epsilon(I) &=
\begin{cases}
P_{\text{unsaturated}} - P, & I > I_{sat} \\
0, & \text{otherwise}
\end{cases}
\label{eq:SFB}
\end{align}
\end{subequations}

In \eqref{eq:AF}, $f(V)$ has three operational regions that are selected based on grid connection standards \cite{ref19}, \cite{ref23}. The first region is the normal operating region when the magnitude of the terminal voltage of the inverter is $V > 0.9$~p.u. In this case, $f(V)=1$; the effective power reference $P^{\ast}$ and droop gain $m_p$ therefore remain unchanged. The second region is the low-voltage ride-through (LVRT) region ($0.5 < V \leq 0.9$~p.u.), where $f(V)$ decreases in proportion to the magnitude of $V$; the effective power reference $P^{\ast}$ and droop gain $m_p$ is therefore reduced accordingly. The third region is the severe voltage dip region ($V \leq 0.5$~p.u.), in which the maximum transferable power is reduced to half of its rated value. In this case, assuming the inverter is injecting maximum power, the power-angle curve is barely above the pre-fault reference. This small deceleration margin causes more severe angle acceleration throughout the severe voltage dip duration and makes loss of synchronism unavoidable. To prevent this, $f(V)$ is set to zero for $V \leq 0.5$~p.u., which fully suppresses active power injection.

The second piecewise function $\epsilon(I)$ in \eqref{eq:SFB} is a feedback term equal to the difference between unsaturated and saturated power. When the magnitude of the inverter current remains within the rated limit ($I \leq I_{sat}$), $\epsilon(I) = 0$. Once the current limiter is activated ($I > I_{sat}$), $\epsilon(I)$ becomes positive and increases with the severity of current saturation. This feedback term $\epsilon(I)$ is to compensate for this current saturation effect. By feeding $\epsilon(I)$ back into \eqref{eq:TTGFM}, the frequency is regulated based on the unsaturated power rather than the saturated power. That eliminates the mismatch of power command during current saturation. These two terms $f(V)$ and $\epsilon(I)$ manipulates the internal reference angle generation of the GFM inverter to act as a virtual braking mechanism. This virtual braking mechanism enhances synchronization stability of the GFM inverters during voltage and frequency dips. Note that TTGFM control does not require a PLL, is independent of the voltage control loop implementation, and needs no extra parameter to tune, supervisory signals, or additional sensors.

\section{Simulation-Based Characterization}
\label{S3}

To evaluate the performance of the TTGFM control, a full-order EMT simulation-based characterization is performed on a GFM inverter connected to a GDE model. GDE model is proposed in \cite{ref24} that represents rest of the grid dynamics considering parasitic interaction between voltage and frequency. The system architecture is shown in Fig.~\ref{fig:Sys_TTGFM}. GFM inverter is interfaced to the GDE with an LCL filter equipped with cascaded voltage and current controllers and a circular current limiter. Synchronization between the inverter and the GDE is achieved through a droop-based PSL, where $L_{\mathrm{eq}}$ and $R_{\mathrm{eq}}$ are the equivalent line inductance and resistance for the rest of the grid. Equivalent inertia of GDE is $H_{\mathrm{eq}}$ and damping is $K_d$ that captures the frequency dynamics of the system. The system  parameters are shown in Table~\ref{tab:System_parameter}.

\begin{table}[h]
\centering
\caption{System parameters}
\label{tab:System_parameter}
\resizebox{\linewidth}{!}{%
\begin{tabular}{lcc|lcc}
\toprule
Parameter & Value & Unit & Parameter & Value & Unit \\
\midrule
\multicolumn{6}{c}{\textbf{Grid Parameters}} \\
\midrule
$S_{\mathrm{grid}}$ & 45    & MW               & $f_0$             & 50    & Hz \\
$L_g$               & 0.026 & p.u.             & $\omega_0$        & $2\pi 50$ & $\mathrm{rad/s}$ \\
$R_g$                & 0.002 & p.u.            & $H_{\mathrm{eq}}$ & 5.08  & s  \\
$l$                  & 2     & km               & $K_d$             & 20    & p.u. \\
\midrule
\multicolumn{6}{c}{\textbf{Inverter Parameters}} \\
\midrule
$S_{\mathrm{vsc}}$   & 5     & MW               & $f_0$               & 50    & Hz \\
$L_{1f}$             & 0.025 & p.u.             & $k_p^{\mathrm{cc}}$ & 0.726 & p.u. \\
$L_{2f}$             & 0.048 & p.u.             & $k_i^{\mathrm{cc}}$ & 0.065 & p.u. \\
$R_{1f}$             & 0.006 & p.u.             & $k_p^{\mathrm{vc}}$ & 0.073 & p.u. \\
$R_{2f}$             & 0.006 & p.u.             & $k_i^{\mathrm{vc}}$ & 0.006 & p.u. \\
$C_f$                & 0.146 & p.u.             & $m_p$               & 5     & \% \\
$\tau_p$             & 33    & $\mathrm{rad/s}$ & $m_q$               & 10    & \% \\
$\tau_q$             & 20    & $\mathrm{rad/s}$ & $I_{\max}$          & 1.2   & p.u. \\
\bottomrule
\end{tabular}%
}
\end{table}

\begin{figure}
    \centering
    \includegraphics[width=0.74\linewidth]{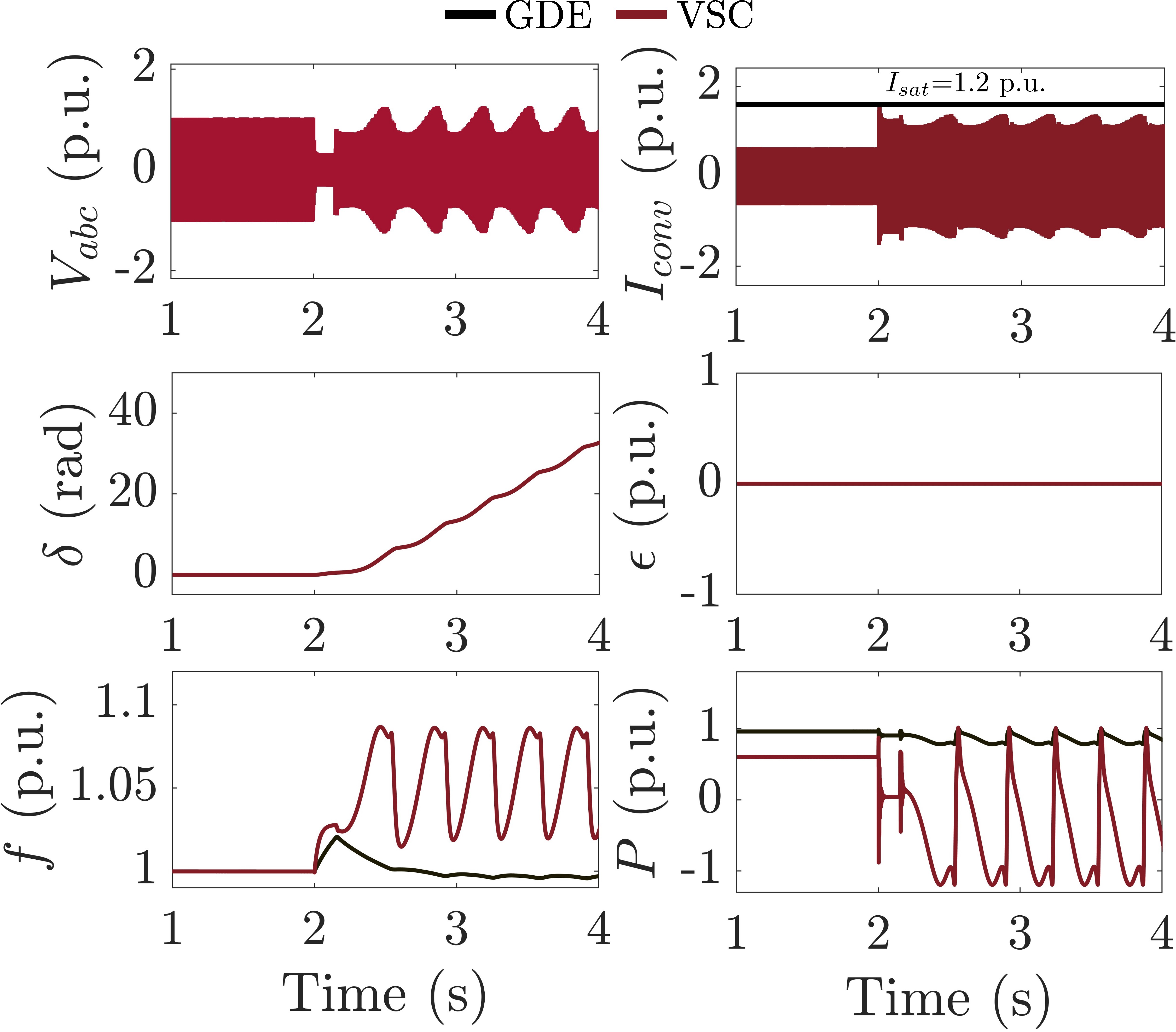}
    \caption{Simulation result for voltage dip with conventional control.}
    \label{fig:Conv100ms}
\end{figure}

\begin{figure}
    \centering
    \includegraphics[width=0.74\linewidth]{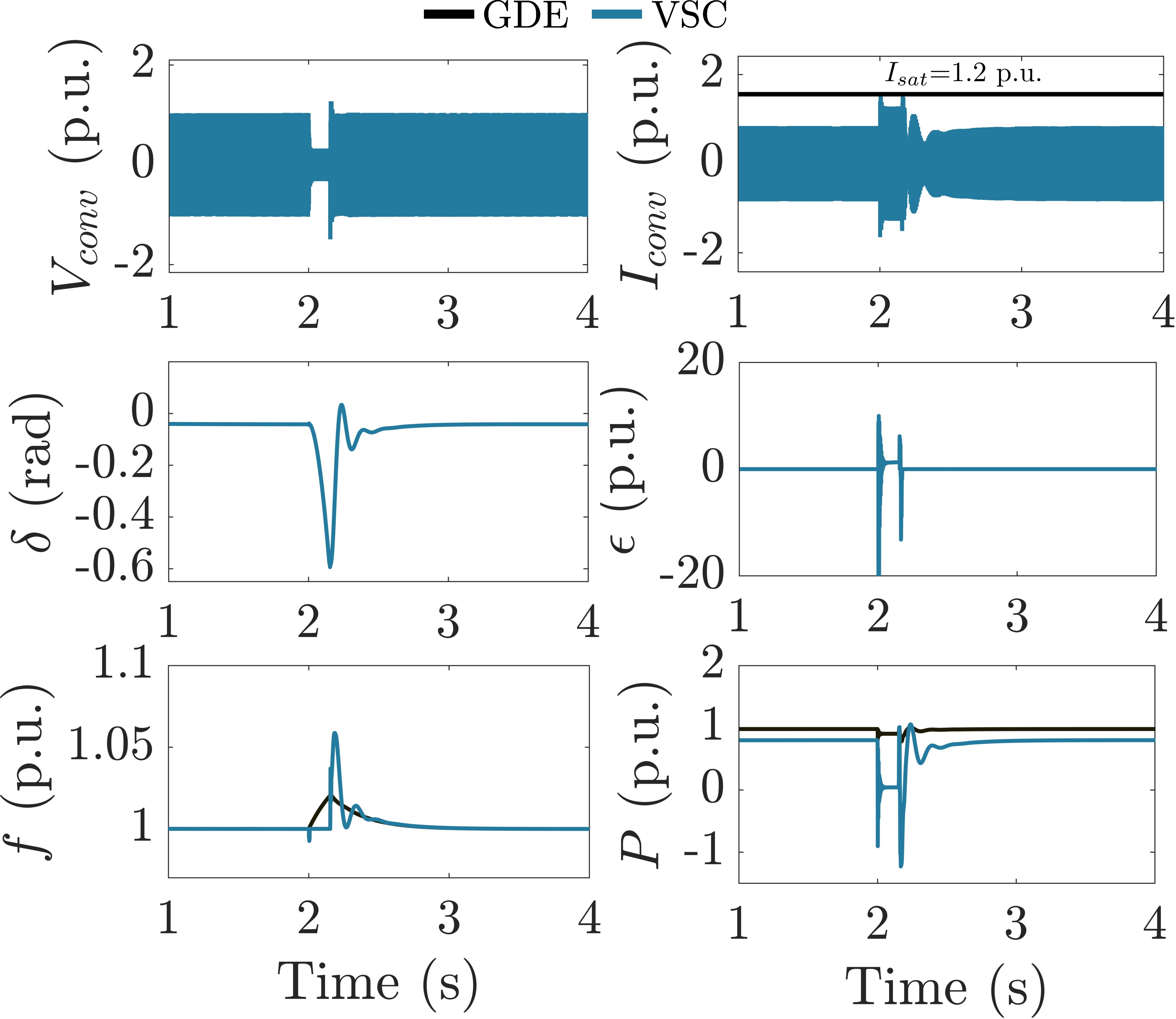}
    \caption{Simulation result for voltage dip with proposed control.}
    \label{fig:Prop100ms}
\end{figure}

\subsection{Voltage Dips}

To evaluate the performance, a voltage dip of $0.3$ pu for a duration of $100$ ms (lower than that lower limit of the fault clearing time; see \cite{ref23}) is applied at $t=2$ s with active power reference $P{^*}$ = 0.6 p.u. The response of TTGFM control is compared against the conventional droop-based GFM control. For the case with conventional control, voltage drop drives the inverter to loss of synchronization as shown in Fig. \ref{fig:Conv100ms}. During voltage dips, GFM inverter activates the current limiter at $I_{\mathrm{sat}}= 1.2$  p.u. for overcurrent protection. So, the angle difference $\delta$ begins to increase immediately after the fault and it increases monotonically, reaching over $30$~rad without settling to any equilibrium. This is because of the error in the droop controller as the inverter is still injecting active power as per the predefined reference. This unbounded acceleration causes inverter frequency $f_{\mathrm{VSC}}$ to deviate from the global frequency of the system $f_{\mathrm{global}}$ rather than converging back to nominal frequency after the disturbance is cleared. So, GFM inverter loses synchronization which causes loss of synchronism for whole system. Note that this behavior of loss of synchronization and instability of the whole system is not visible with the infinite bus system (see, e.g., \cite{ref13}, \cite{ref14}, \cite{ref20}, \cite{ref21}, \cite{ref22}).

In contrast, with the proposed TTGFM control as seen in Fig. \ref{fig:Prop100ms}, the angle difference $\delta$ deviates by less than $-0.6$~rad during voltage dip. Rather than accelerating indefinitely after $100$ ms, $\delta$ returns to its pre-fault value. This is because of the piecewise function defined in \eqref{eq:AF} reduces the effective power reference $P^*$ and droop gain $m_p$. Also, as there is a difference between unsaturated and saturated power, $\epsilon(I)$ becomes positive. Together, these two terms act as the virtual braking mechanism as described in Section~\ref{S2} limits angle acceleration. When the disturbance is cleared, GFM inverter resynchronizes with the rest of the grid, thanks to the proposed TTGFM control. This validates that TTGFM control extends the synchronization stability margin during voltage dips compared to the conventional control.

\subsection{Frequency Dips}
\label{S2.2}

\begin{figure}
    \centering
    \includegraphics[width=0.74\linewidth]{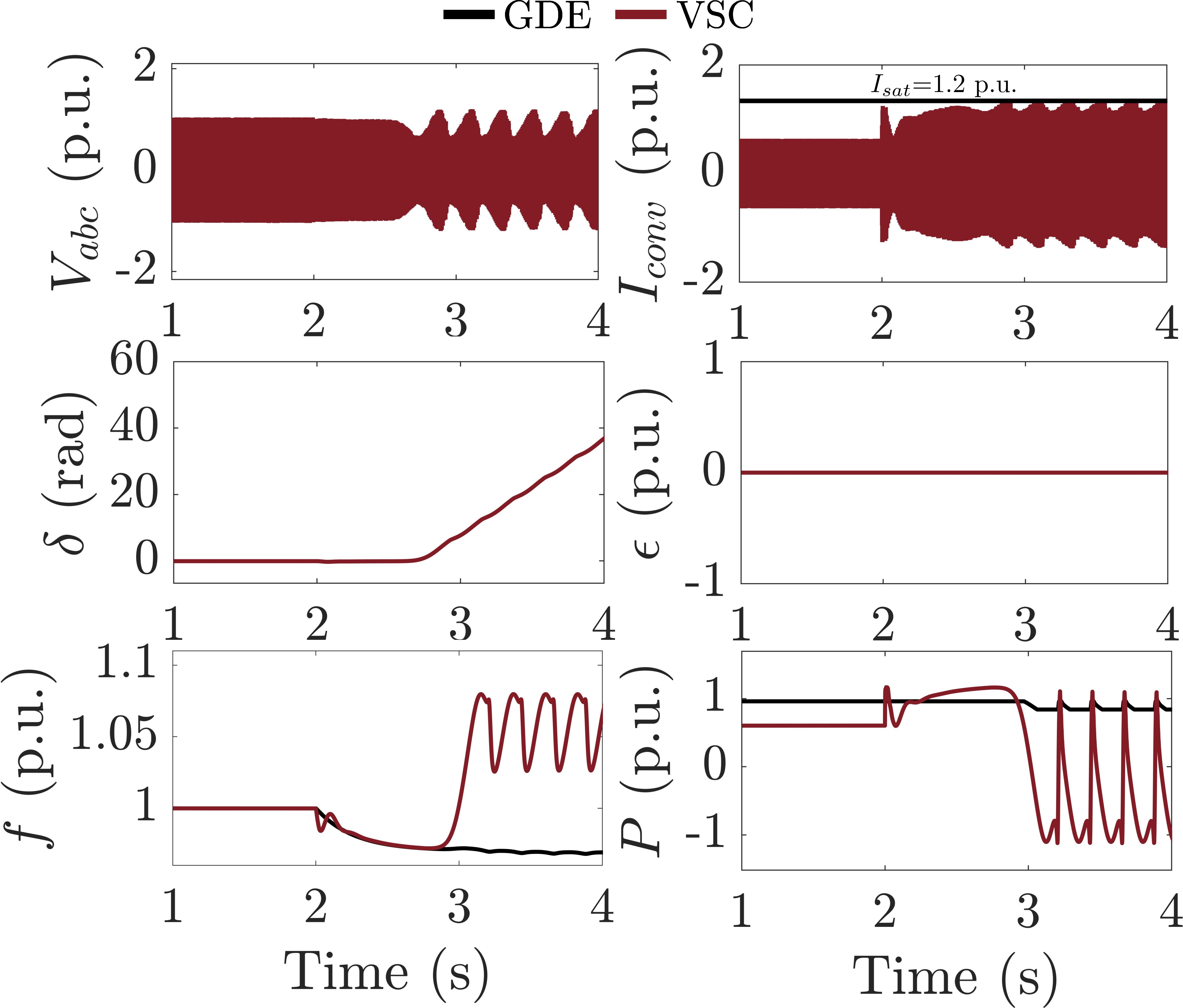}
    \caption{Simulation result for frequency dip with conventional control.}
    \label{fig:Conv6per}
\end{figure}

\begin{figure}
    \centering
    \includegraphics[width=0.74\linewidth]{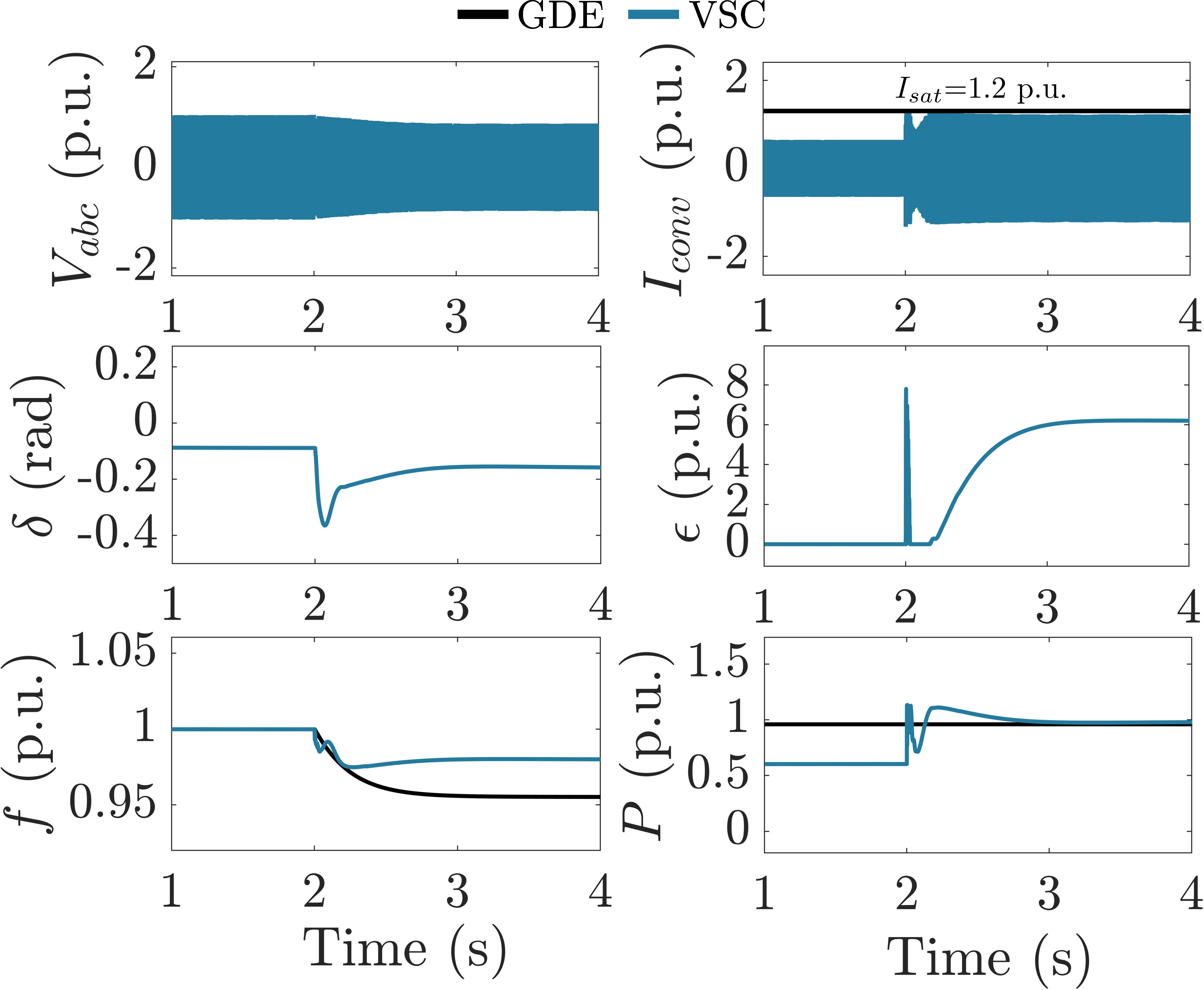}
    \caption{Simulation result for frequency dip with proposed control.}
    \label{fig:Prop6per}
\end{figure}

To evaluate the performance under frequency dip, a positive load disturbance (increase in load) of $6\%$ of the inverter rating is applied at $t=2$~s with inverter operating at $P^*=0.6$~pu. This load disturbance is sufficient to trigger the current limiter. Here, the response of the TTGFM control is also compared against the conventional droop-based GFM control. Unlike voltage dip, the angle is not accelerating immediately upon load disturbance at $t=2$~s; instead, the loss of synchronization develops gradually over time. 

For the case of conventional control, the load disturbance drives the inverter into loss of synchronism as shown in Fig.~\ref{fig:Conv6per}. Following the disturbance, $P_{\mathrm{VSC}}$ initially increases to balance the load demand. However, as the current limiter activates the angle difference $\delta$ begins to accelerate around $t=2.7$~s and increases monotonically thereafter, reaching over $60$~rad by $t=5$~s without settling to any equilibrium. This is because the droop controller continues to push power to balance the load disturbance. This unbounded acceleration causes $f_{\mathrm{VSC}}$ to deviate from $f_{\mathrm{global}}$. So, GFM inverter loses synchronization which causes loss of synchronism for whole system.

In contrast, with the proposed TTGFM control as seen in Fig.~\ref{fig:Prop6per}, the angle difference $\delta$ deviates by less than $0.3$~rad from its pre-disturbance value before settling to a new steady-state value. Rather than accelerating indefinitely, $\delta$ converges within approximately $1$~s of disturbance. This is because once the current limiter is activated, difference between the unsaturated and saturated power increases i.e., second piecewise function $\epsilon(I)$ becomes positive. The terminal voltage of the GFM inverter also decreases from steady-state during the load disturbance. It is because of the fact that the output voltage of GFM inverter is coupled to the internal angle $\theta$ through the $V\text{-}\theta$ relationship. As the load disturbance causes $\theta$ to decrease, the voltage magnitude decreases correspondingly (see \cite{ref20}). So, $f(V)$ reduces the effective power reference $P^*$ and droop gain $m_p$. As a result, the virtual braking mechanism is activated and resists angle acceleration to enhance synchronization stability margin. Both $f_{\mathrm{VSC}}$ and $P_{\mathrm{VSC}}$ settles to new steady-state values respectively confirming that the GFM inverter remains synchronized with the grid, thanks to the proposed TTGFM control. This validates that the TTGFM control extends the synchronization stability margin during frequency dip compared to the conventional control.

\section{Benchmarking Against State-of-the-Art Enhanced Synchronization Methods}
\label{S4}

\begin{figure}
    \centering
    \includegraphics[width=0.74\linewidth]{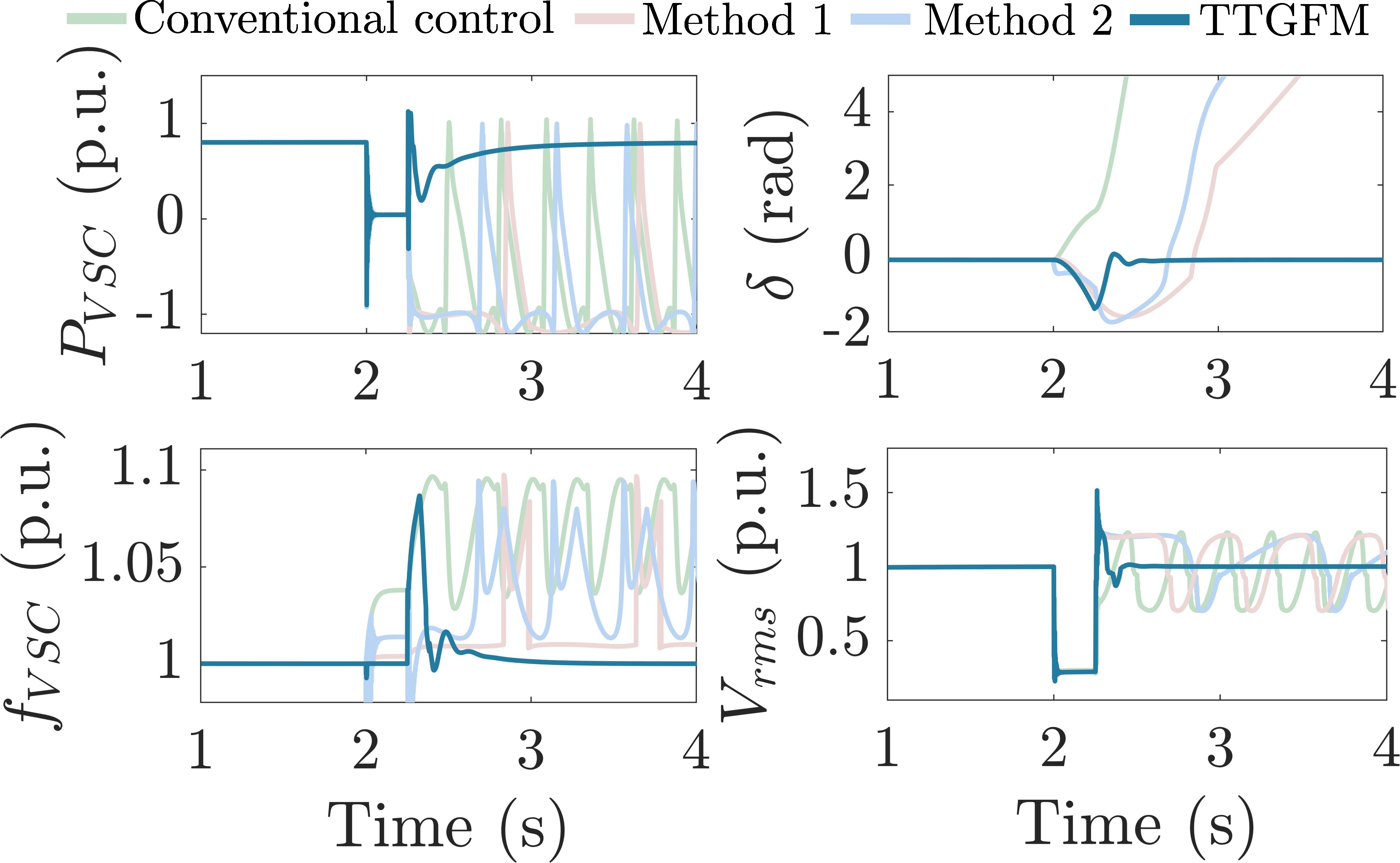}
    \caption{Benchmarking results for voltage dip.}
    \label{fig:BMVD}
\end{figure}

\begin{figure}
    \centering
    \includegraphics[width=0.74\linewidth]{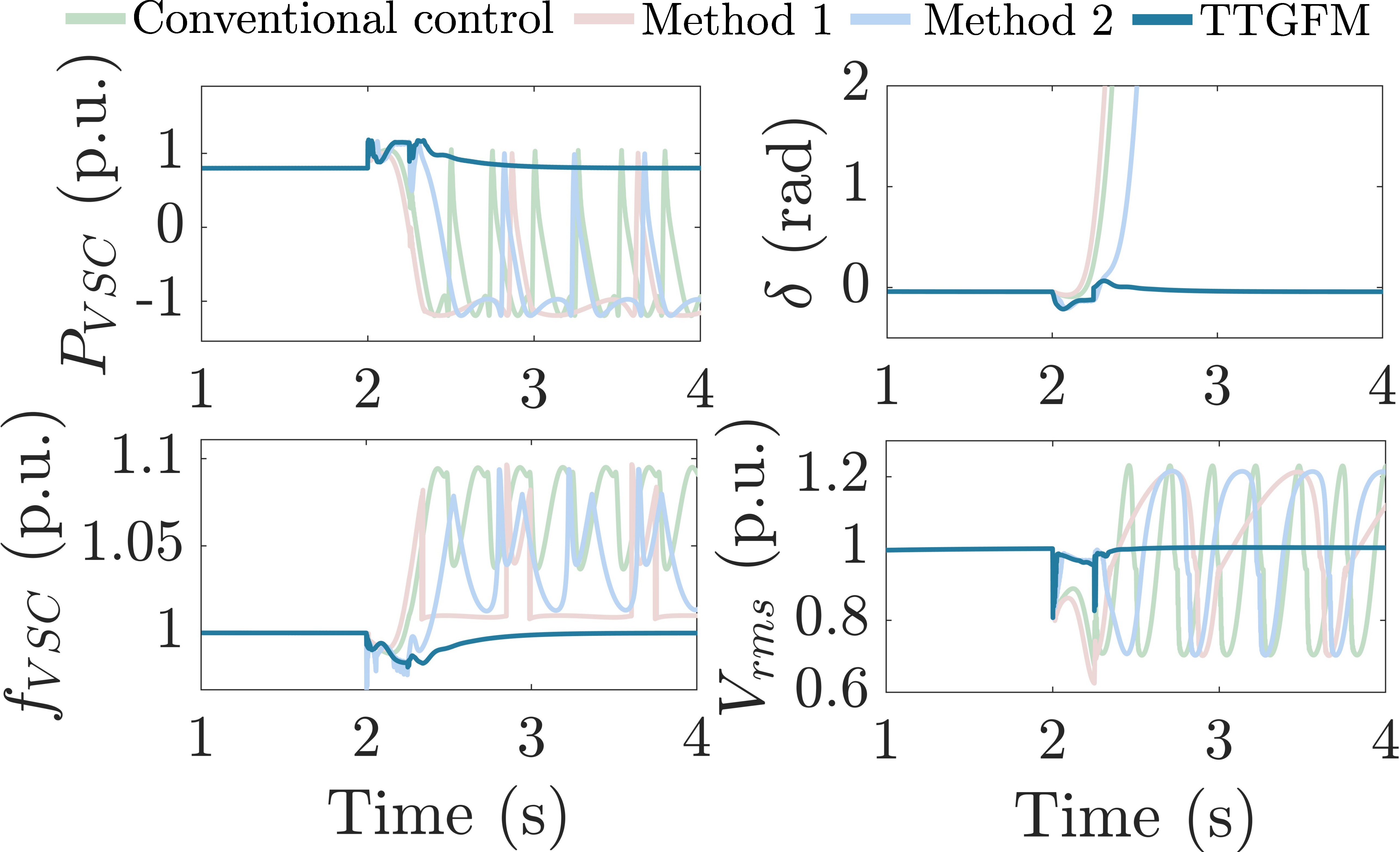}
    \caption{Benchmarking results for frequency dip.}
    \label{fig:BMFD}
\end{figure}

To validate the effectiveness, the proposed TTGFM control is benchmarked against two other enhanced synchronization schemes proposed in the literature. In method 1 \cite{ref11}: droop gain $m_{p}$ is reduced to one-tenth of its nominal value once the output current exceeds the defined threshold. Note that, in this method this reduction to one-tenth is a fixed gain that is multiplied with the nominal droop gain. In Method 2 \cite{ref17}: active power reference is reduced by using a proportional gain once the output current exceeds the defined threshold. This proportional gain is tuned to enhance the performance as per the disturbance. Both methods introduce a modification in the PSL to enhance performance under disturbances. To evaluate the performance, each method is subjected to disturbances specified as per ENTSO-E requirements \cite{ref19}, \cite{ref23}: a) grid voltage dip of $0.3$~p.u. for $250$~ms, and b) load disturbance of $7.31\%$ (of inverter rating)  for $250$~ms that corresponds to a rate of change of frequency (RoCoF) of $-4$~Hz/s. In this case the setpoints of the inverter is increased to $P^\ast = 0.8$\,pu.

For the disturbance of grid voltage dip, it is observed in Fig.~\ref{fig:BMVD} that the conventional control, method 1, and method 2 show loss of synchronization. This is because the methods  are not enough to limit the angle acceleration. As a result, internal angle $\delta$ increases monotonically, and after disturbance clearance the inverter fails to resynchronize. In contrast, proposed TTGFM control can limit the angle acceleration and resynchronize to the pre-fault equilibrium after the disturbance is cleared. 

For load disturbance, similar phenomenon is observed in Fig.~\ref{fig:BMFD}. In conventional control, Method 1 and Method 2, synchronism is lost. Internal angle $\delta$ increases monotonically, and the inverter fails to resynchronize once the disturbance is cleared. But the proposed TTGFM control can limit the angle acceleration and resynchronize to the pre-fault equilibrium after the disturbance is cleared. This confirms grid code compliance of the TTGFM control. Note that the transient actions introduced by the piecewise function in \eqref{eq:PF} is triggered only during disturbances and, thereby do not affect the steady-state operation. 

In summary, these benchmarking results confirm that synchronism is not maintained by conventional control, Method 1, or Method 2 for both disturbances. Whereas, TTGFM control maintains synchronism and restores stable operation after both of the disturbances are cleared and compatible with the connection requirement standards. Table~\ref{tab:BM} summarizes these findings: cross ($\times$) indicates loss of synchronism, and check mark (\checkmark) indicates that synchronism is maintained.

\begin{table}[htbp]
\centering
\caption{Summary of Benchmarking Results}
\label{tab:BM}
\begin{tabular}{lcc}
\hline
\textbf{Methods} & \textbf{Voltage dip} & \textbf{Load disturbance} \\
\hline
Conventional control    & $\times$ & $\times$ \\
Method 1                & $\times$ & $\times$ \\
Method 2                & $\times$ & $\times$ \\
TTGFM control & \checkmark & \checkmark \\
\hline
\end{tabular}
\end{table}

\begin{figure*}[t!]
\centering
\includegraphics[width=0.74\linewidth]{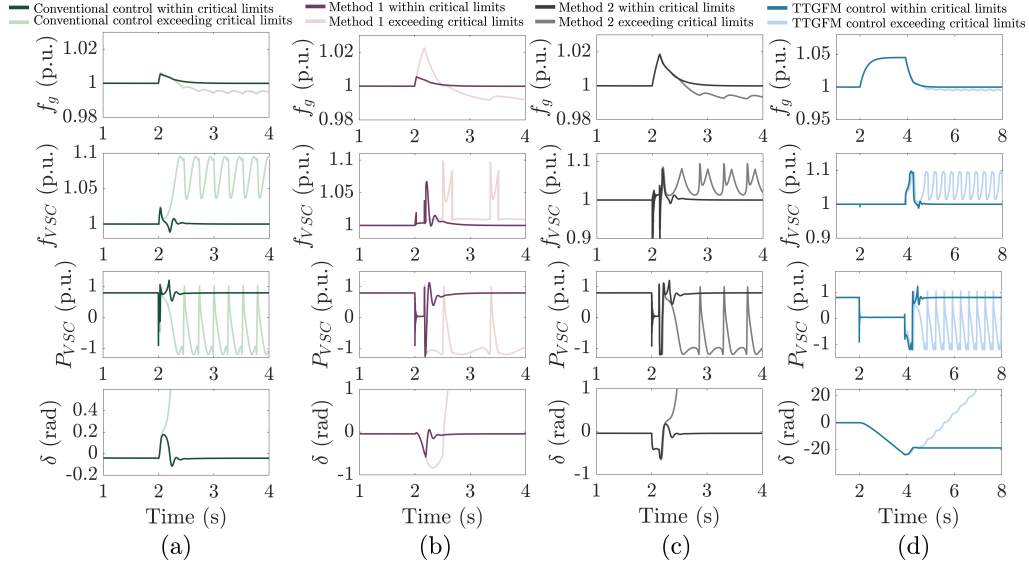}
\caption{Critical limit for voltage dip (a) Conventional control, (b) Method 1 (c) Method 2 (d) TTGFM control.}
\label{fig:CL_BF}
\end{figure*}

\section{Analytical Framework}
\label{S5}

\subsection{Preliminaries}
\label{S4A}

The analytical modeling here corresponds to the system architecture of Fig.~\ref{fig:Sys_TTGFM}. GFM inverter is interfaced to the GDE through an LCL filter that has an inverter-side inductor ($L_{1f}$, $R_{1f}$), a filter capacitor ($C_f$), and a grid-side inductor ($L_{2f}$, $R_{2f}$). The inverter output is connected at the point of common coupling (PCC) while the rest of the grid is represented by the GDE that has equivalent inertia $H_{eq}$ and equivalent line impedance $Z_{eq}$. GDE model represents mixed and fast dynamics of voltage and frequency as they are coupled together. All quantities are expressed in $dq0$ reference frame rotating at the synchronized frequency $50$ Hz with the alignment in $d$-axis that is expressed in \eqref{eq:Vgdq}.

\begin{equation}
V_{gd} = V_g, \qquad V_{gq} = 0
\label{eq:Vgdq}
\end{equation}

Here, $V_g$ denotes the GDE voltage magnitude. In this frame, the inverter voltage is expressed as:

\begin{equation}
v_{d} = V\cos\delta, \qquad v_{q} = V\sin\delta
\label{eq:Vidq}
\end{equation}

Where $V$ is the inverter terminal voltage magnitude and $\delta$ denotes the phase angle difference between the inverter and the GDE. As power flows from the inverter toward the GDE, the expression is given in \eqref{eq:PQ}.

\begin{equation}
p = \frac{3}{2}v_{d}\,i_{2d}, \qquad 
q = -\frac{3}{2}v_{d}\,i_{2q}
\label{eq:PQ}
\end{equation}

Here $i_{2d}$ and $i_{2q}$ denote the grid-side $dq$-axis currents of the LCL filter. The dynamic equations of LCL filter is given by \eqref{eq:LCL_dynamics}.

\begin{equation}
\begin{aligned}
\dot{i}_{1d} &= -\frac{R_1}{L_1}i_{1d} + \omega_0 i_{1q} - \frac{v_{cfd}}{L_1} + \frac{1}{L_1}v_{1d} \\
\dot{i}_{1q} &= -\frac{R_1}{L_1}i_{1q} - \omega_0 i_{1d} - \frac{v_{cfq}}{L_1} + \frac{1}{L_1}v_{1q} \\
\dot{i}_{2d} &= -\frac{R_2}{L_2}i_{2d} + \omega_0 i_{2q} + \frac{v_{cfd}}{L_2} - \frac{1}{L_2}v_d \\
\dot{i}_{2q} &= -\frac{R_2}{L_2}i_{2q} - \omega_0 i_{2d} + \frac{v_{cfq}}{L_2} - \frac{1}{L_2}v_q \\
\dot{v}_{cfd} &= \frac{1}{C_f}i_{1d} - \frac{1}{C_f}i_{2d} + \omega_0 v_{cfq} \\
\dot{v}_{cfq} &= \frac{1}{C_f}i_{1q} - \frac{1}{C_f}i_{2q} - \omega_0 v_{cfd}
\end{aligned}
\label{eq:LCL_dynamics}
\end{equation}

To protect the semiconductor switches, a circular current limiter is used. When $I$ exceeds the rated threshold $I_{\max}$, the limiter scales down both $i_d$ and $i_q$ by the same ratio by preserving the phase angle of the current. The scaling factor is given by \eqref{eq:limiter}.

\begin{equation}
\sigma =
\begin{cases}
1, & I \le I_{\max} \\[4pt]
\dfrac{I_{\max}}{I}, & I > I_{\max}
\end{cases}
\label{eq:limiter}
\end{equation}

In PSL, droop controller with a low-pass filter in active power measurement is considered. Active power measured after the low-pass filter is $P_f$. The expression of PSL is denoted by \eqref{eq:droop}.

\begin{equation}
\omega = \omega^* + m_p\bigl(P^*-P_f\bigr), \qquad \tau_p\dot{P}_f+P_f=p
\label{eq:droop}
\end{equation}

Here $\omega^*$ is the nominal frequency reference, $P^*$ the active power setpoint, $m_p$ is the droop gain, $\tau_p$ the time constant of the low-pass filter and $p$ is the active power as defined in \eqref{eq:PQ}. The internal angle dynamics is given by \eqref{eq:delta_dot}.

\begin{equation}
\dot{\delta} = \omega-\omega_g
\label{eq:delta_dot}
\end{equation}

Eliminating $P_f$ through differentiation of \eqref{eq:delta_dot} and substituting \eqref{eq:droop} and \eqref{eq:PQ} gives the second-order angle dynamics in \eqref{eq:delta_ode}.

\begin{equation}
\begin{aligned}
\ddot{\delta} =\;
&\underbrace{-\tfrac{1}{\tau_p}\dot{\delta}}_{\text{damping}}
+\underbrace{\tfrac{1}{\tau_p}\bigl(\omega^*-\omega_g\bigr)}_{\text{frequency error}}
+\underbrace{\tfrac{m_p}{\tau_p}\bigl(P^*-P\bigr)}_{\text{droop controller}}
-\underbrace{\dot\omega_g}_{\substack{\text{frequency} \\ \text{dynamics}}}
\end{aligned}
\label{eq:delta_ode}
\end{equation}

Reactive power is also controlled by voltage droop and $\tau_q$ is the time constant of the low-pass filter. The dynamics of the voltage droop is expressed in \eqref{eq:volt_ode}.

\begin{equation}
\dot{V} = \tfrac{1}{\tau_q}\bigl(V^*-V\bigr) + \tfrac{m_q}{\tau_q}\bigl(Q^*-Q_f\bigr)
\label{eq:volt_ode}
\end{equation}

Rather than assuming a constant frequency system as in a quasi-static model (see, e.g., \cite{ref25}), GDE is represented with its own frequency dynamics. Frequency dynamics of the GDE is given by $\omega_g$ in \eqref{eq:grid_freq}, and the GDE angle $\theta_g$ is obtained by integrating $\omega_g$.

\begin{equation}
\dot\omega_g = \tfrac{1}{2H_{eq}}\bigl(P+P_g-P_L-K_d(\omega_g-\omega_0)\bigr)
\label{eq:grid_freq}
\end{equation}

$H_{eq}$ is the equivalent inertia that is calculated from \cite{ref26}, $P_g$ denotes the active power setpoints generated by the rest of the grid, $P_L$ the global load demand, and $\omega_0$ the nominal grid frequency. The three-phase voltage dynamics of GDE is expressed as \eqref{eq:GDE_voltage}.

\begin{equation}
\begin{aligned}
V_{GDE}^a &= V_g\sin\theta_g \\
V_{GDE}^b &= V_g\sin\!\left(\theta_g - \tfrac{2\pi}{3}\right) \\
V_{GDE}^c &= V_g\sin\!\left(\theta_g + \tfrac{2\pi}{3}\right)
\end{aligned}
\label{eq:GDE_voltage}
\end{equation}

\begin{figure*}[t!]
\centering
\includegraphics[width=0.74\linewidth]{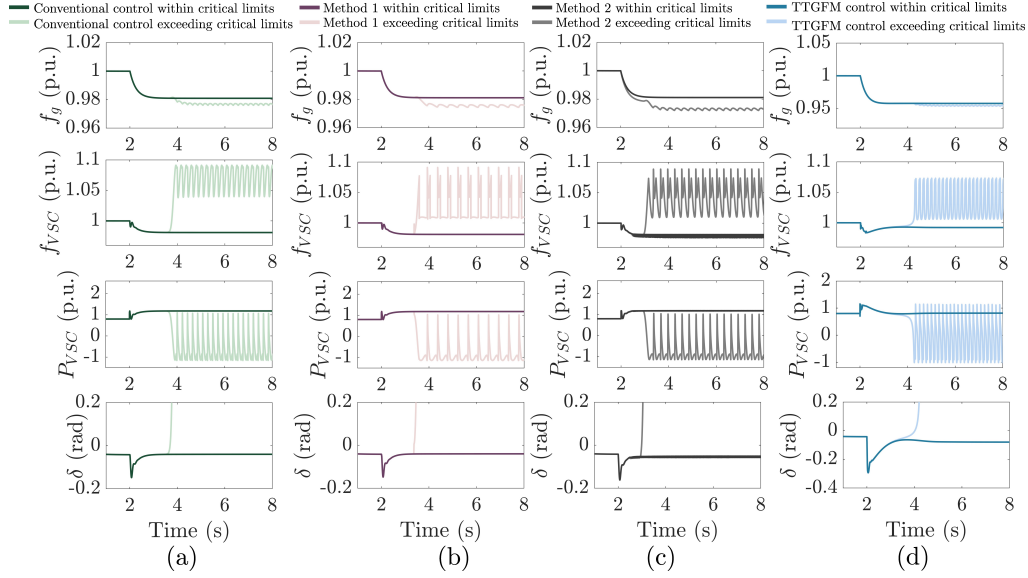}
\caption{Critical limit for frequency dip (a) Conventional control, (b) Method 1 (c) Method 2 (d) TTGFM control.}
\label{fig:CL_load}
\end{figure*}

Here, $V_g$ is the GDE voltage magnitude defined in \eqref{eq:Vgdq}. The nonlinear model is obtained by collecting all the state variables from \eqref{eq:Vidq}-\eqref{eq:grid_freq}. This framework augmented with the current limiter dynamics shown in \eqref{eq:limiter} forms the basis for the synchronization instability analysis in the following sections.

\subsection{Synchronization Instability Mechanism}
\label{S4B}

The second-order angle dynamics $\ddot\delta$ given by \eqref{eq:delta_ode} is analyzed for the two disturbances that is discussed in Sections~\ref{S3}.

\subsubsection{Voltage Dip}

A voltage dip corresponds to a reduction of voltage magnitude. This reduction generally depends on fault impedance $Z_f$ and equivalent line impedance $Z_{eq}$. PCC voltage in this case is expressed in \eqref{eq:pcc_voltage}.

\begin{equation}
v_d = V_g \frac{Z_f}{Z_{eq}+Z_f}, \qquad v_q = 0
\label{eq:pcc_voltage}
\end{equation}

Substituting $v_{d}$ from \eqref{eq:pcc_voltage} and grid frequency dynamics $\dot\omega_g$ from \eqref{eq:grid_freq} in the second-order angle dynamics equation in \eqref{eq:delta_ode}, gives the angle dynamics expression during voltage dip in \eqref{eq:ddelta_fault_evo_zeq}.

\begin{equation}
\begin{aligned}
\ddot{\delta} = &-\frac{\dot\delta}{\tau_p} + \frac{1}{\tau_p}\left(\omega^*-\omega_g\right) + \frac{m_p}{\tau_p}\left(P^* - \frac{3}{2}\,\frac{V_g Z_f}{Z_{eq}+Z_f}\,i_{2d}\right) \\
&- \frac{1}{2H_{eq}}\left(\frac{3}{2}\,\frac{V_g Z_f}{Z_{eq}+Z_f}\,i_{2d} + P_g - P_L - K_d(\omega_g-\omega_0)\right)
\end{aligned}
\label{eq:ddelta_fault_evo_zeq}
\end{equation}

During bolted fault PCC voltage collapses and the factor $Z_f/(Z_{eq}+Z_f)$ in \eqref{eq:ddelta_fault_evo_zeq} vanishes, such that $v_d\to 0$. Consequently, active power $P$ delivered by the GFM inverter also collapses to zero. Note that $P_g$ does not collapse as it is a constant setpoint from the rest of the grid. This mismatch between $P_g$ and $P_L$ governs the frequency dynamics of the system. During a bolted fault as $v_d=0$ in \eqref{eq:ddelta_fault_evo_zeq} that simplifies the second-order angle dynamics expression to \eqref{eq:ddelta_fault_evo}.

\begin{equation}
\ddot{\delta} = -\frac{\dot\delta}{\tau_p} + \bigl(\omega^*-\omega_g\bigr)\!\left(\frac{1}{\tau_p}-\frac{K_d}{2H_{eq}}\right) + \frac{m_p P^*}{\tau_p} + \frac{P_L-P_g}{2H_{eq}}
\label{eq:ddelta_fault_evo}
\end{equation}

The term $m_pP^*/\tau_p$ is the input of the conventional droop controller and remains unchanged during voltage dip and the term $(P_L-P_g)/2H_{eq}$ is the exogenous signal that is the dynamic response of $\dot\omega_g$. Because $P=0$ and input of the droop controller $m_pP^*/\tau_p$ in \eqref{eq:ddelta_fault_evo} is positive, $\ddot\delta$ keeps accelerating during voltage dips. So, if $\delta$ surpasses a critical angle $\delta_{\mathrm{cr}}$ before the disturbance is cleared, the GFM inverter loses synchronism with the grid. Beyond this angle, no stable equilibrium exists since $\ddot{\delta}$ increases monotonically.

Nevertheless, PCC voltage in  \eqref{eq:pcc_voltage} is scaled by $\frac{Z_f}{Z_{eq}+Z_f}$ for non-zero $Z_f$. A larger $Z_{eq}$ (weaker grid) reduces this factor that causes more severe voltage dip and thus a larger reduction in $P$ for a given time. This results in faster angle acceleration and small critical clearing time (CCT) compared to strong grid (small $Z_{eq}$). Effect of $Z_{eq}$ is discussed in details in section \ref{S6}.

\subsubsection{Frequency dip}

Frequency dip corresponds to a sudden load disturbance of magnitude $\Delta P_L>0$, which introduces a negative RoCoF i.e., frequency dip. So, the mismatch between $P_L$ and $P_g$ in \eqref{eq:grid_freq} adds an accelerating force to the grid frequency dynamics. At the same time, the droop controller tries to compensate for the load disturbance by commanding additional current; this yields a restoring term proportional to the current deviation $\Delta i_{d}=i_{d}(t)-i_{d}^0$, where $i_{d}^0$ is the current just before the load step. Substituting the expression of $p$ from \eqref{eq:PQ} and the grid frequency dynamics from \eqref{eq:grid_freq} into \eqref{eq:delta_ode} gives the full expression of the angle dynamics in \eqref{eq:ddelta_load_evo}.

\begin{equation}
\begin{aligned}
\ddot{\delta} &= -\frac{\dot\delta}{\tau_p} + (\omega_g-\omega^*)\!\left(\frac{K_d}{2H_{eq}}-\frac{1}{\tau_p}\right) + \frac{P_{L}+\Delta P_L-P_{g}}{2H_{eq}} \\
&\quad - \left(\frac{m_p}{\tau_p}+\frac{1}{2H_{eq}}\right)\!v_g\Delta i_{d}(t)
\end{aligned}
\label{eq:ddelta_load_evo}
\end{equation}

The term $\Delta P_L$ is the exogenous signal introduced by the load step and the magnitude is positive throughout the disturbance. In the last term, droop gain $m_p$ and grid inertia $H_{eq}$ act as a restoring force. But this is valid only if the inverter delivers the required current ($\Delta i_{d}>0$). Once the current limiter is activated $i_{d}$ becomes saturated. This means, the restoring term is saturated and $\ddot\delta$ keeps accelerating during frequency dip. So, if $\delta$ surpasses a critical angle $\delta_{\mathrm{cr}}$ before the disturbance is cleared, the GFM inverter loses synchronism with the grid. Beyond this angle, no stable equilibrium exists since $\ddot{\delta}$ increases monotonically.

\subsection{Angle Dynamics with TTGFM Control}
\label{S4C}

\subsubsection{Modified Angle Dynamics}

Substituting the TTGFM control law from \eqref{eq:TTGFM} and following the same derivation procedure of Section~\ref{S4A} gives the second‑order angle dynamics ($\ddot{\delta}_{\mathrm{TT}}$) of TTGFM control in \eqref{eq:delta_ddot_TT_eps}.

\begin{equation}
\begin{array}{c}
\ddot{\delta}_{\mathrm{TT}} =\;
-\dfrac{1}{\tau_p}\dot{\delta}
+\dfrac{1}{\tau_p}\bigl(\omega^*-\omega_g\bigr)
+\dfrac{m_p f^2(V)}{\tau_p}\Bigl(P^*-P-\epsilon(I)\Bigr) \\[6pt]
-\dot\omega_g
\end{array}
\label{eq:delta_ddot_TT_eps}
\end{equation}

Using $\epsilon(I) = P_{\mathrm{unconstrained}}-P$ from \eqref{eq:SFB}, and subtracting and expanding \eqref{eq:delta_ddot_TT_eps} from \eqref{eq:delta_ode} isolates the virtual braking terms that is shown in \eqref{eq:TT_reduction}.  

\begin{equation}
\begin{aligned}
\ddot{\delta}_{\mathrm{TT}} = \ddot{\delta}
&- \underbrace{\frac{m_p}{\tau_p}(P^*-P)\bigl[1-f^2(V)\bigr]}_{\text{virtual braking}}- \underbrace{\frac{m_p f^2(V)\,\epsilon(I)}{\tau_p}}_{\text{virtual braking}}
\end{aligned}
\label{eq:TT_reduction}
\end{equation}

Both braking terms are positive because $P^*-P\ge0$, $f(V)\in[0,1]$, and $\epsilon(I)\ge0$. Hence $\ddot{\delta}_{\mathrm{TT}}\le\ddot{\delta}$: TTGFM control never accelerates faster than the conventional droop law and brakes to slow down acceleration when either of the terms is present.

\subsubsection{Voltage Dip}

During voltage dip, $f(V)$ reduces $P^*$ and $m_p$ as the terminal voltage of the GFM inverter dips. Also, $\epsilon(I)$ is positive as the current limiter activates. So, angle dynamics for voltage dip by TTGFM control is expressed as \eqref{eq:ddelta_fault_evo_TT}.

\begin{equation}
\begin{aligned}
\ddot{\delta}_{\mathrm{TT}} = -\frac{\dot\delta}{\tau_p} + \bigl(\omega^*-\omega_g\bigr)\!\left(\frac{1}{\tau_p}-\frac{K_d}{2H}\right) \\
+ \underbrace{\frac{m_p f^2(V)\bigl(P^*-\epsilon(I)\bigr)}{\tau_p}}_{\text{virtual braking}} + \frac{P_L-P_g}{2H_{eq}}
\end{aligned}
\label{eq:ddelta_fault_evo_TT}
\end{equation}

The term $m_p f^2(V)\bigl(P^*-\epsilon(I)\bigr)/\tau_p$ replaces constant droop input $m_pP^*/\tau_p$ from \eqref{eq:ddelta_fault_evo} that acts a virtual braking mechanism. As voltage dip increases with severity, $f(V)<1$ slows down the acceleration and when the current limiter is active, $\epsilon(I)$ becomes more positive, which slows down acceleration further. Unlike conventional droop control, proposed TTGFM control reduces the angle acceleration and maintains internal angle $\delta$ below $\delta_{\mathrm{cr}}$ for a given time. 

\subsubsection{Frequency Dip}
During a sudden load disturbance, the current limiter triggers if the inverter reaches the current capability. As a result, $\epsilon(I)$ becomes positive and increases according to the severity. Also, as the terminal voltage of the GFM inverter also drops as explained in \ref{S2.2}, $f(V)<1$ reduces effective $P^*$ and $m_p$. The angle dynamics for a frequency dip under TTGFM control is expressed in \eqref{eq:ddelta_load_evo_TT}.

\begin{equation}
\begin{aligned}
\ddot{\delta}_{\mathrm{TT}} &= -\frac{\dot\delta}{\tau_p} + (\omega_g-\omega^*)\!\left(\frac{K_d}{2H_{eq}}-\frac{1}{\tau_p}\right) \\
&\quad + \frac{P_{L}+\Delta P_L-P_{g}}{2H_{eq}} - \left(\frac{m_p}{\tau_p}+\frac{1}{2H_{eq}}\right)\!V_g\Delta i_{d}(t) \\
&\quad - \underbrace{\frac{m_p f^2(V)\,\epsilon(I)}{\tau_p}}_{\text{virtual braking}}
\end{aligned}
\label{eq:ddelta_load_evo_TT}
\end{equation}

The first four terms are identical to the conventional angle dynamics equation for frequency dip in \eqref{eq:ddelta_load_evo}. The extra terms $\epsilon(I)$ and $-m_p f^2(V)\epsilon(I)/\tau_p$ is introduced because of the piecewise functions and act as a virtual braking mechanism. Unlike conventional droop control, the proposed TTGFM control here also reduces the angle acceleration and maintains the internal angle $\delta$ below $\delta_{\mathrm{cr}}$ for a given time. So, proposed TTGFM control extends the synchronization stability margin of the GFM inverter during voltage and frequency dips. 

\subsection{Critical Limit}

It can be concluded from the previous section that TTGFM control improves the synchronization stability margin of the GFM inverter during voltage and frequency dips. Nevertheless, synchronization may still be lost in both scenarios, because the second term in \eqref{eq:ddelta_fault_evo_TT} and \eqref{eq:ddelta_load_evo_TT}, which is \((\omega_g-\omega^*)\!\left(\frac{K_d}{2H_{eq}}-\frac{1}{\tau_p}\right)\) can drive \(\delta\) toward \(\delta_{\mathrm{cr}}\). This term is constant and independent of the disturbance type, except for its dependence on \(\omega_g\), and is not compensated by TTGFM control. Meanwhile, \(\omega_g\) varies dynamically with the frequency deviation. Since \((\omega_g-\omega^*)\) remains positive, a longer disturbance duration continues to  push \(\delta\) toward \(\delta_{\mathrm{cr}}\) that eventually causing loss of synchronization irrespective of the virtual braking mechanism. 

\begin{table}[t!]
\footnotesize
\setlength{\tabcolsep}{2pt}
\centering
\caption{Summary of critical limits}
\label{tab:critical_limits_by_metric}
\resizebox{\columnwidth}{!}{%
\begin{tabular}{clcccccc}
\toprule
\textbf{Parameters} & \textbf{Line Length} & \textbf{SCR} & \textbf{Conventional} & \textbf{Method 1 \cite{ref11}} & \textbf{Method 2 \cite{ref17}} & \textbf{TTGFM} \\
\midrule
{CCT}
 & 2 km & 3.8    & 27 ms & 169 ms & 130 ms & 1.91 s \\
 & 8 km & 0.95 & 20 ms & 136 ms & 130 ms & 1.38 s  \\
\midrule
Critical load\textsuperscript{1}
 & 2 km & 3.8    & 3.83\% & 3.78\% & 4.40\% & 11.70\% \\
RoCoF\textsuperscript{2}
 & 2 km & 3.8    & 2.09 Hz/s & 2.07 Hz/s & 2.46 Hz/s & 6.39 Hz/s \\
\bottomrule
\end{tabular}%
}
\begin{minipage}{\columnwidth}
\footnotesize \textsuperscript{1}Percentages are expressed relative to the base power of VSC. \\
\textsuperscript{2}RoCoF corresponding to the critical load for each method.
\label{tab:CCL}
\end{minipage}
\end{table}

\begin{figure}
    \centering
\includegraphics[width=0.77\linewidth]{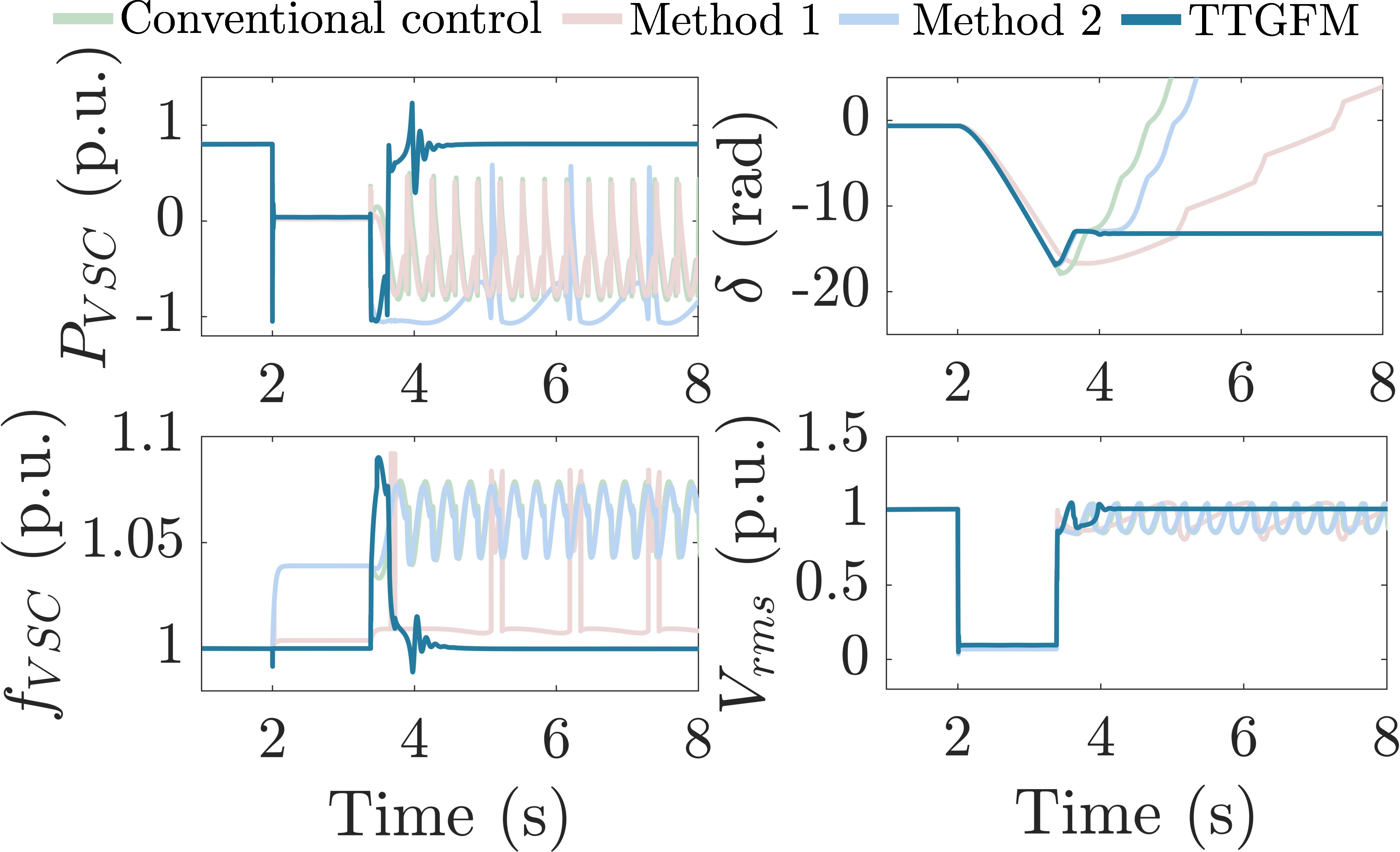}
    \caption{Critical limit for voltage dip of line length=8 km.}
    \label{fig:LLV}
\end{figure}

\begin{figure*}[t!]
\centering
\includegraphics[width=0.74\linewidth]{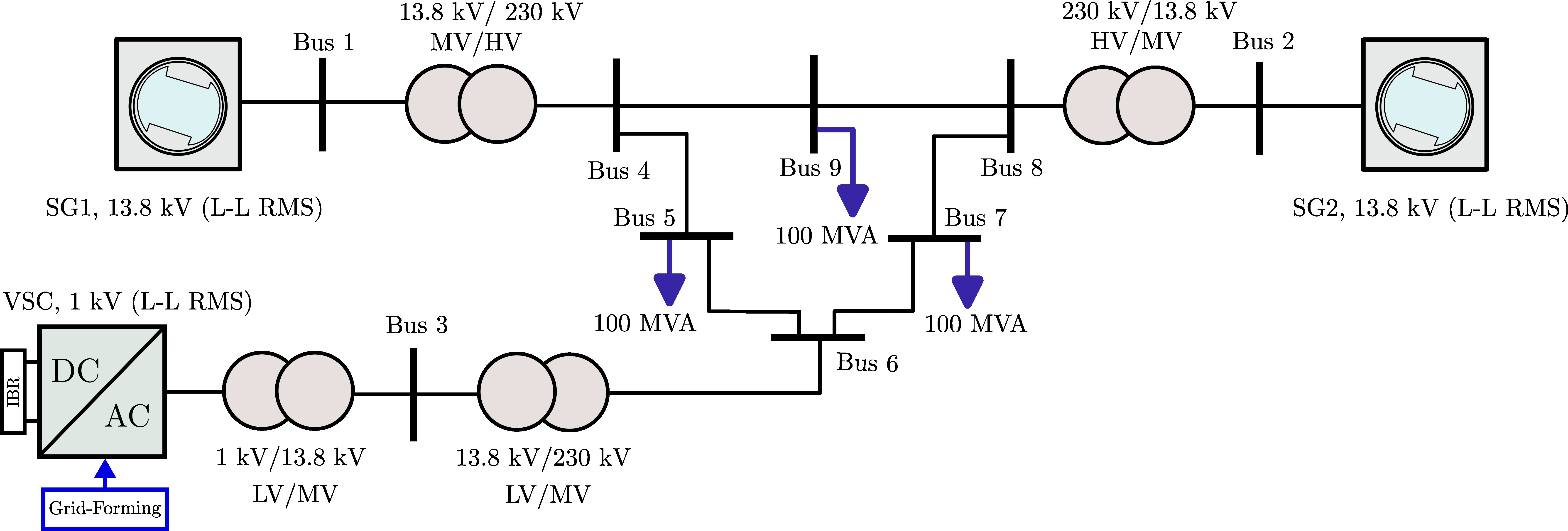}
\caption{Modified IEEE 9-bus system.}
\label{fig:MIII9bus}
\end{figure*}

\begin{table}[t!]
\footnotesize
\setlength{\tabcolsep}{3pt}
\centering
\caption{Comparison of GDE Model with IEEE 9-Bus System}
\label{tab:GDE_validation}
\resizebox{\columnwidth}{!}{%
\begin{tabular}{lccccccc}
\toprule
& \multicolumn{2}{c}{\textbf{\shortstack{Short-circuit\\current (p.u.)}}} & \multicolumn{2}{c}{\textbf{\shortstack{Equivalent\\impedance (p.u.)}}} & \textbf{\shortstack{Equivalent\\inertia}} & \textbf{CCT} \\
\cmidrule(lr){2-3} \cmidrule(lr){4-5}
\textbf{Model} & \textbf{Analytical} & \textbf{Simulation} & \textbf{Analytical} & \textbf{Simulation} & & \\
\midrule
IEEE 9-bus & 5.151 & 5.20 & 0.194 & 0.192 & 5.08~s & 29~ms \\
GDE        & 5.076 & 5.50 & 0.196 & 0.181 & 5.08~s & 27~ms \\
\bottomrule
\end{tabular}%
}
\end{table}

\begin{figure}
    \centering    \includegraphics[width=0.87\linewidth]{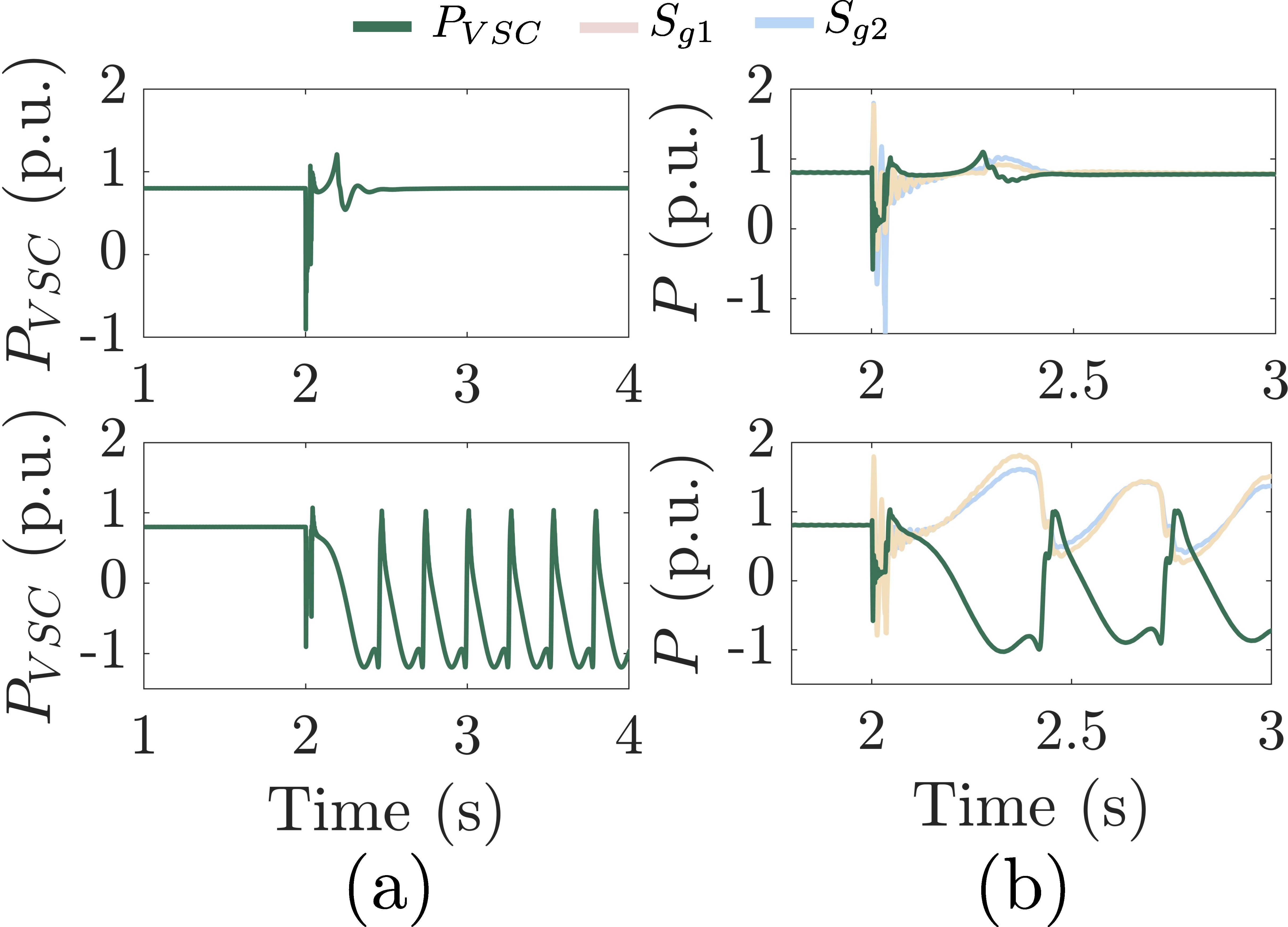}
    \caption{CCT for a voltage dip.  (a) GDE (b) IEEE 9-bus. Top: within critical limits; bottom: exceeding critical limits.}
    \label{fig:GDE_g}
\end{figure}

In other words, synchronization can still be lost if $(\omega_g-\omega^*)$ remains nonzero for long enough as it is not compensated by TTGFM control. Fig.~\ref{fig:CL_BF} and~\ref{fig:CL_load} show the critical limits for voltage and frequency dips respectively. Note that, $p$ is labeled as $P_{VSC}$ in these figures. The critical limit of TTGFM control is again compared with conventional droop control, Method~1 \cite{ref11}, and Method~2 \cite{ref17}. TTGFM control achieves the highest critical limit, thanks to the virtual braking mechanism. Even a disturbance exceeding the critical limit by just 0.1\% causes loss of synchronism in all of these methods as the GFM inverter can no longer return to a stable equilibrium. Table~\ref{tab:CCL} summarizes the critical limits.   

\subsection{Effect of Critical Limit for Line Impedance Variation}

To evaluate the effect of the line impedance $Z_{eq}$ on the critical limits, the line connecting the GFM inverter to the GDE is changed. The length is varied from $L=2\,\mathrm{km}$ (SCR $=3.8$) to $L=8\,\mathrm{km}$ (SCR $=0.95$), while all other parameters remain unchanged. A shorter line reduces the line impedance and effectively keeps the GFM inverter closer to the grid. Whereas, increasing the line length to 8~km increases the line impedance (see, \eqref{eq:pcc_voltage}) and weakens the grid strength. Fig.~\ref{fig:LLV} shows the critical limit for a line length of $L=8\,\mathrm{km}$, and Table~\ref{tab:critical_limits_by_metric} summarizes the resulting critical limits for both line lengths. As shown in Fig.~\ref{fig:LLV}, the critical limit for CCT is 1.38~s for $L=8\,\mathrm{km}$, smaller than the CCT obtained for $L=2\,\mathrm{km}$. Also, it is seen from \eqref{eq:ddelta_fault_evo_zeq} that if $Z_{eq}$ increases, the angle acceleration is faster, so the CCT is smaller. On the other hand, smaller $Z_{eq}$ gives a slower angle acceleration and a larger CCT. This confirms that a longer line reduces transient stability margin compared to shorter lines. Nevertheless for both line lengths, TTGFM control retains substantially larger synchronization stability margin indicating robustness to variations in grid impedance.

\section{Generalization of TTGFM Control to Benchmark Systems}
\label{S6}

The purpose of this section is to demonstrate that the proposed TTGFM control is not limited to the GDE model, but is applicable to any standard benchmark or real systems. This generalization holds as long as rest of the grid can be accurately represented by an equivalent GDE model. In this article, IEEE 9-bus system is used as a representative benchmark. It is first reduced to its equivalent GDE parameters ($H_{eq}$ and $Z_{eq}$), and the dynamic response of this equivalent is then compared against the modified IEEE 9-bus system. 

Fig.~\ref{fig:MIII9bus} shows the modified IEEE 9-bus system considered in this article, where the GFM inverter is connected at bus 6. The equivalent inertia $H_{eq}$ is obtained via a curve-fitting method \cite{ref26}. The equivalent impedance $Z_{eq}$ of the rest of the grid which is seen from bus 6 can be obtained using two approaches: 1) Analytically, by using Kirchhoff's laws, as the branch impedances are known, and 2) by simulation, applying a short circuit at bus 6 and measuring the line current. Both methods are applied to the modified IEEE 9-bus system, and the obtained $H_{eq}$ and $Z_{eq}$ are implemented in the GDE model. The results are summarized in Table~\ref{tab:GDE_validation} (see \cite{ref17} for IEEE 9-bus system parameters). 

Fig. \ref{fig:GDE_g} shows that the CCT obtained during voltage dip from the GDE model closely matches to the modified IEEE 9-bus system. This confirms that the GDE model can replicate the transient behavior of this standard benchmark system. This validation also concludes that the proposed analytical framework with TTGFM control in section \ref{S4} is not restricted to a particular system but is valid for any standard benchmarked or real system if $H_{eq}$ and $Z_{eq}$ are correctly extracted. Thus, the analytical framework and TTGFM control can therefore be extended to other systems also  (e.g., IEEE 14-bus and IEEE 39-bus).

\section{Conclusion}
\label{S7}

This article proposed a TTGFM control to enhance the synchronization stability of GFM inverters under voltage and frequency dips. By feeding back the terminal voltage and the difference between unsaturated and saturated active power into the PSL, the proposed method acts as a virtual braking mechanism that limits internal angle acceleration whenever the current limiter is triggered. This transient triggering action is activated only during disturbances and does not affect steady-state operation. High-fidelity EMT simulation results showed that conventional droop control loses synchronism for voltage and frequency dips, whereas TTGFM control limits the angle acceleration and resynchronizes to pre-disturbance equilibrium once the disturbance is cleared. Benchmarking against two state-of-the-art synchronization stability enhancement schemes further confirmed that TTGFM control maintains synchronism under both disturbance types simultaneously satisfying grid code requirements including line impedance variations. An analytical framework was also developed to characterize the synchronization instability mechanism using GDE model by incorporating grid dynamics. The analysis showed that although TTGFM control limits the internal angle acceleration but yet synchronism can still be lost if the grid frequency deviation continues for long enough. Also, TTGFM control is not restricted to a single configuration but applicable to any standard benchmark system as concluded through generalization of GDE. Finally, critical limit showed that TTGFM control has larger synchronization stability margins compared to conventional control and the benchmarked methods.

See \cite{ref1,ref2,ref3,ref4,ref5}

\end{document}